# Bottom quark production cross section at HERA-B

N. Kidonakis and J. Smith

*Institute for Theoretical Physics, State University of New York at Stony Brook, Stony Brook, NY 11794-3840, USA*

### Abstract

The cross section for bottom quark production is calculated for the HERA-B experiment. We consider both the order $\alpha_s^3$ cross section and the resummation of soft gluon corrections in all orders of QCD perturbation theory.

# 1    Introduction

The calculation of production cross sections for heavy particles in QCD is made by invoking the factorization theorem [1] and expanding the contributions to the amplitude in powers of the coupling constant $\alpha_s(\mu^2)$. Recent investigations have shown that near threshold there can be large logarithms in the perturbation expansion which have to be resummed to make more reliable theoretical predictions. The application of these ideas to fixed-target Drell-Yan production has been the subject of many papers over the past few years [2]. The same ideas on resummation were applied to the calculation of the top-quark cross section and differential distributions at the Fermilab Tevatron in [3-5]. What is relevant in these reactions is the existence of a class of logarithms of the type $(\ln(1-z))^i/(1-z)$, where $i$ is the order of the perturbation expansion, and where one must integrate over the variable $z$ up to a limit $z = 1$. These terms are not actually singular at $z = 1$ due to the presence of terms in $\delta(1-z)$. However the remainder can be quite large. In general one writes such terms as "plus" distributions, which are then convoluted with regular test functions (the parton densities).

In this paper we examine the production of $b$-quarks in a situation where the presence of these large logarithms is of importance, namely in a fixed-target experiment to be performed in the HERA ring at DESY. This actual experiment has the name HERA-B [6, 7] and involves colliding the circulating proton beam against a stationary copper wire in the beam pipe. The nominal beam energy of the protons is 820 GeV, so that the square root of the center-of-mass (c.m.) energy is $\sqrt{S} = 39.2$ GeV. Taking the $b$-quark mass as $m_b = 4.75 \, \text{GeV}/c^2$ then the ratio of $m_b/\sqrt{S} \approx 1/8$. If we choose the renormalization scale in the running coupling constant as $m_b$ then $\alpha_s(m_b^2) \approx 0.2$ so $\alpha_s(m_b^2)\ln(\sqrt{S}/m_b) \approx 0.4$, indicating that perturbation theory should be reliable.

In perturbation theory with a hard scale we can use the standard expression for the order-by order cross section in QCD, namely

$$\sigma(S, m^2) = \int_{\frac{4m^2}{S}}^{1} dx_1 \int_{\frac{4m^2}{Sx_1}}^{1} dx_2 \sum_{ij} f_i(x_1, \mu^2) f_j(x_2, \mu^2) \sigma_{ij}(s = x_1 x_2 S, m^2, \mu^2),$$

(1.1)

where the $f_i(x, \mu^2)$ are the parton densities at the mass factorization scale $\mu^2$ and the $\sigma_{ij}$ are the partonic cross sections. The numerical results for the hadronic cross sections depend on the parton densities, which involve the choice of $\mu^2$; the choice of the running coupling constant, which involves the renormalization scale (also normally chosen to be $\mu^2$); and the choice for the actual mass of the $b$-quark. In lowest order (LO) or Born approximation the actual numbers for the cross section show a large sensitivity to these parameters. The next-to-leading order (NLO) results follow from the work of the two groups [8] and [9, 10]. However, even the NLO results do not completely fix the cross section. There is still a sensitivity to our lack of knowledge of even higher terms in the QCD expansion, which can be demonstrated by varying the scale choice up and down by factors of two. In general it is impossible to make more precise predictions for heavy quark production given the absence of a calculation in next-to-next-to-leading order (NNLO).

In the threshold region one can improve on the NLO results. In this region one finds that there are large logarithms of the type mentioned above which arise from the soft-plus-virtual (S+V) terms in the perturbation expansion. These logarithms can be resummed to all orders in perturbation theory. We will see later that the gluon-gluon channel is the dominant channel for the production of $b$-quarks near threshold in a fixed-target $pp$ experiment. In the next section we give some results for the parton-parton cross section. Section 3 contains the analysis of the hadron-hadron cross section which is relevant for the HERA-B experiment. We give results in NLO and after resummation. Finally in Section 4 we give our conclusions.

## 2  Results for parton-parton reactions

The partonic processes that we examine are

$$i(k_1) + j(k_2) \rightarrow Q(p_1) + \bar{Q}(p_2), \tag{2.1}$$

where $i, j = g, g$ or $i, j = q, \bar{q}$ and $Q, \bar{Q}$ are heavy quarks $(c, b, t)$. The square of the parton-parton c.m. energy is $s = (k_1 + k_2)^2$.

We begin with heavy quark production in the $q\bar{q}$ channel. The Born cross section in this channel is given by

$$\sigma_{q\bar{q}}^{(0)}(s, m^2) = \frac{2\pi}{3} \alpha_s^2(\mu^2) K_{q\bar{q}} N C_F \frac{1}{s} \beta \left(1 + \frac{2m^2}{s}\right), \tag{2.2}$$

where $C_F = (N^2 - 1)/(2N)$ is the Casimir invariant for the fundamental representation of $SU(N)$, $K_{q\bar{q}} = N^{-2}$ is a color average factor, $m$ is the heavy quark mass, $\mu$ denotes the renormalization scale, and $\beta = \sqrt{1 - 4m^2/s}$. Also $N = 3$ for the $SU(3)$ color group in QCD. The threshold behavior ($s \to 4m^2$) of this expression is given by

$$\sigma_{q\bar{q},\,\mathrm{thres}}^{(0)}(s, m^2) = \pi\alpha_s^2(\mu^2)K_{q\bar{q}}NC_F\frac{1}{s}\beta. \qquad (2.3)$$

Complete analytic results are not available for the NLO cross section as some integrals are too complicated to do by hand. However in [10] analytic results are given for the soft-plus-virtual contributions to the cross section, and for the approximation to the cross section near threshold. Simple formulae which yield reasonable approximations to the exact $O(\alpha_s^3)$ results have been constructed in [11].

The analysis of the contributions to the gluon-gluon channel in NLO is much more complicated. There are three Born diagrams and different color structures. The exact Born term in the $gg$ channel is

$$
\begin{aligned}
\sigma_{gg}^{(0)}(s, m^2) \;=\; & 4\pi\alpha_s^2(\mu^2)K_{gg}NC_F\frac{1}{s}\bigg\{ C_F\bigg[ -\bigg(1 + \frac{4m^2}{s}\bigg)\beta \\
& + \bigg(1 + \frac{4m^2}{s} - \frac{8m^4}{s^2}\bigg)\ln\frac{1+\beta}{1-\beta}\bigg] \\
& + C_A\bigg[ -\bigg(\frac{1}{3} + \frac{5}{3}\frac{m^2}{s}\bigg)\beta + \frac{4m^4}{s^2}\ln\frac{1+\beta}{1-\beta}\bigg]\bigg\}, \qquad (2.4)
\end{aligned}
$$

where $K_{gg} = (N^2 - 1)^{-2}$ is a color average factor and $C_A = N$ is the Casimir invariant for the adjoint representation of $SU(N)$. The threshold behavior ($s \to 4m^2$) of this expression is given by

$$\sigma_{gg,\,\mathrm{thres}}^{(0)}(s, m^2) = \pi\alpha_s^2(\mu^2)K_{gg}\frac{1}{s}NC_F[4C_F - C_A]\beta. \qquad (2.5)$$

Again, the complete NLO expression for the cross section in the $gg$ channel is unavailable but analytic results are given for the S+V terms in [9]. These were used in [11] to analyze the magnitude of the cross section near threshold.

In [3] an approximation was given for the NLO soft-plus-virtual (S+V) contributions and the analogy with the Drell-Yan process was exploited to resum them to all orders of perturbation theory. We are discussing partonic reactions of the type $i(k_1) + j(k_2) \to Q(p_1) + \bar{Q}(p_2) + g(k_3)$, and we introduce the kinematic variables $t_1 = (k_2 - p_2)^2 - m^2$, $u_1 = (k_1 - p_2)^2 - m^2$, and $s_4 = s + t_1 + u_1$. The variable $s_4$ depends on the four-momentum of the extra partons emitted in the reaction and is zero for elastic scattering. The first-order S+V result in the $\overline{\text{MS}}$ scheme is

$$
\begin{aligned}
s^2 \frac{d^2\sigma_{ij}^{(1)}(s,t_1,u_1)}{dt_1 du_1} &= \sigma_{ij}^B(s,t_1,u_1)\frac{2C_{ij}}{\pi}\alpha_s(\mu^2) \\
&\times \left[ \frac{1}{s_4}\left(2\ln\frac{s_4}{m^2} + \ln\frac{m^2}{\mu^2}\right)\theta(s_4 - \Delta) \right. \\
&\left. + \left(\ln^2\frac{\Delta}{m^2} + \ln\frac{\Delta}{m^2}\ln\frac{m^2}{\mu^2}\right)\delta(s_4) \right]
\end{aligned}
\tag{2.6}
$$

where $\Delta$ is a small parameter used to distinguish between the soft $(s_4 < \Delta)$ and the hard $(s_4 > \Delta)$ regions in phase space. Here we define $C_{q\bar{q}} = C_F$, $C_{gg} = C_A$,

$$
\sigma_{q\bar{q}}^B(s,t_1,u_1) = \pi\alpha_s^2(\mu^2)K_{q\bar{q}}NC_F\left[\frac{t_1^2 + u_1^2}{s^2} + \frac{2m^2}{s}\right],
\tag{2.7}
$$

and

$$
\begin{aligned}
\sigma_{gg}^B(s,t_1,u_1) &= 2\pi\alpha_s^2(\mu^2)K_{gg}NC_F\left[C_F - C_A\frac{t_1 u_1}{s^2}\right] \\
&\times \left[\frac{t_1}{u_1} + \frac{u_1}{t_1} + \frac{4m^2 s}{t_1 u_1}\left(1 - \frac{m^2 s}{t_1 u_1}\right)\right].
\end{aligned}
\tag{2.8}
$$

We can define an analogous result for the $q\bar{q}$ channel in the DIS scheme (but not for the $gg$ channel) which is

$$
\begin{aligned}
s^2 \frac{d^2\sigma_{q\bar{q}}^{(1)}}{dt_1 du_1}(s,t_1,u_1) &= \sigma_{q\bar{q}}^B(s,t_1,u_1)\frac{2C_F}{\pi}\alpha_s(\mu^2) \\
&\times \left[ \frac{1}{s_4}\left(\ln\frac{s_4}{m^2} + \ln\frac{m^2}{\mu^2}\right)\theta(s_4 - \Delta) \right. \\
&\left. + \left(\frac{1}{2}\ln^2\frac{\Delta}{m^2} + \ln\frac{\Delta}{m^2}\ln\frac{m^2}{\mu^2}\right)\delta(s_4) \right].
\end{aligned}
\tag{2.9}
$$

The resummation of the leading S+V terms has been given in [3]. The result is

$$
s^2 \frac{d^2 \sigma_{ij}^{\mathrm{res}}(s, t_1, u_1)}{dt_1 du_1} = \sigma_{ij}^B(s, t_1, u_1) \left[ \frac{df(s_4/m^2, m^2/\mu^2)}{ds_4} \theta(s_4 - \Delta) \right.
$$
$$
\left. + f(\frac{\Delta}{m^2}, \frac{m^2}{\mu^2}) \delta(s_4) \right], \tag{2.10}
$$

where

$$
f\left( \frac{s_4}{m^2}, \frac{m^2}{\mu^2} \right) = \exp \left[ A \frac{C_{ij}}{\pi} \bar{\alpha}_s \left( \frac{s_4}{m^2}, m^2 \right) \ln^2 \frac{s_4}{m^2} \right] \frac{[s_4/m^2]^\eta}{\Gamma(1+\eta)} \exp(-\eta \gamma_E). \tag{2.11}
$$

Expressions for $A$, $\bar{\alpha}_s$, $\eta$, and $\gamma_E$ are given in [3].

# 3 Results for bottom quark production at HERA-B

In this section we discuss $b$-quark production at HERA-B, and we examine the effects of the resummation procedure discussed in the previous section. Following the notation in [3] the total hadron-hadron cross section in order $\alpha_s^k$ is

$$
\sigma_H^{(k)}(S, m^2) = \sum_{ij} \int_{4m^2/S}^1 d\tau \, \Phi_{ij}(\tau, \mu^2) \, \sigma_{ij}^{(k)}(\tau S, m^2, \mu^2), \tag{3.1}
$$

where $S$ is the square of the hadron-hadron c.m. energy and $i, j$ run over $q, \bar{q}$ and $g$. The parton flux $\Phi_{ij}(\tau, \mu^2)$ is defined via

$$
\Phi_{ij}(\tau, \mu^2) = \int_\tau^1 \frac{dx}{x} H_{ij}(x, \frac{\tau}{x}, \mu^2), \tag{3.2}
$$

and $H_{ij}$ is a product of the scale-dependent parton distribution functions $f_i^h(x, \mu^2)$, where $h$ stands for the hadron which is the source of the parton $i$

$$
H_{ij}(x_1, x_2, \mu^2) = f_i^{h_1}(x_1, \mu^2) f_j^{h_2}(x_2, \mu^2). \tag{3.3}
$$

The mass factorization scale $\mu$ is chosen to be identical with the renormalization scale in the running coupling constant.

In the case of the all-order resummed expression the lower boundary in (3.1) has to be modified according to the condition $s_0 < s - 2ms^{1/2}$, where $s_0$ is defined below (see [3]). Resumming the soft gluon contributions to all orders we obtain

$$\sigma_H^{res}(S, m^2) = \sum_{ij} \int_{\tau_0}^1 d\tau \, \Phi_{ij}(\tau, \mu^2) \, \sigma_{ij}(\tau S, m^2, \mu^2) \,, \qquad (3.4)$$

where $\sigma_{ij}$ is given in (3.24) of [3] and

$$\tau_0 = \frac{[m + (m^2 + s_0)^{1/2}]^2}{S} \,, \qquad (3.5)$$

with $s_0 = m^2(\mu_0^2/\mu^2)^{3/2}$ ($\overline{MS}$ scheme) or $s_0 = m^2(\mu_0^2/\mu^2)$ (DIS scheme). Here $\mu_0$ is the non-perturbative parameter used in [3]. It is used to cut off the resummation since the resummed corrections diverge for small $\mu_0$.

We now specialize to bottom quark production at HERA-B where $\sqrt{S} = 39.2$ GeV. In the presentation of our results for the exact, approximate, and resummed hadronic cross sections we use the MRSD$_-'$ parametrization for the parton distributions [12]. Note that the hadronic results only involve partonic distribution functions at moderate and large $x$, where there is little difference between the various sets of parton densities. We have used the MRSD$_-'$ set 34 as given in PDFLIB [13] in the DIS scheme with the number of active light flavors $n_f = 4$ and the QCD scale $\Lambda_5 = 0.1559$ GeV. We have used the two-loop corrected running coupling constant as given by PDFLIB. Note that we have checked scheme differences in the $q\bar{q}$ channel by using the MRSD$_-'$ set 31 as given in PDFLIB [13] in the $\overline{MS}$ scheme. For the $gg$ channel there is no DIS scheme so we always use the $\overline{MS}$ scheme.

First, we discuss the NLO contributions to bottom quark production at HERA-B using the results in [8-10]. Except when explicitly stated otherwise we will take the factorization scale $\mu = m_b$ where $m_b$ is the $b$-quark mass. Also, throughout the rest of this paper, we will use $m$ and $m_b$ interchangeably. In fig. 1 we show the relative contributions of the $q\bar{q}$ channel in the DIS scheme and the $gg$ channel as a function of the bottom quark mass. We see that the $gg$ contribution is the dominant one, lying between 70% and 80% of the total NLO cross section for the range of bottom mass values given. The $q\bar{q}$ contribution is smaller and makes up most of the remaining cross section. The relative contributions of the $gq$ and the $g\bar{q}$ channels in the DIS scheme are negative and very small. The situation here is the reverse of what is known about top quark production at the Fermilab Tevatron where $q\bar{q}$ is the dominant channel with $gg$ making up the remainder of the cross section, and $gq$ and $g\bar{q}$ making an even smaller relative contribution than is the case for bottom quark production at HERA-B. The reason for this difference between top quark and bottom quark production is that the Tevatron is a $p\bar{p}$ collider while HERA-B is a fixed-target $pp$ experiment. Thus, the parton densities involved are different and since sea quark densities are much smaller than valence quark densities, the $q\bar{q}$ contribution to the hadronic cross section diminishes for a fixed-target $pp$ experiment relative to a $p\bar{p}$ collider for the same partonic cross section.

In fig. 2 we show the $K$ factors for the $q\bar{q}$ and $gg$ channels and for their sum as a function of bottom quark mass. The $K$ factor is defined by $K = (\sigma^{(0)} + \sigma^{(1)}\mid_{\text{exact}})/\sigma^{(0)}$, where $\sigma^{(0)}$ is the Born term and $\sigma^{(1)}\mid_{\text{exact}}$ is the exact first order correction. We notice that all $K$ factors are large. This is to be expected due to the new dynamical mechanisms which arise in NLO. The figure shows that higher order effects are more important for the $gg$ channel than for $q\bar{q}$. The $K$ factor for the sum of the two channels is also quite large. However, the $K$ factor for the total is slightly lower since we also include the negative contributions of the $qg$ and $\bar{q}g$ channels.

These large corrections come predominantly from the threshold region for bottom quark production where it has been shown that initial state gluon bremsstrahlung (ISGB) is responsible for the large cross section at NLO [11]. This can easily be seen in fig. 3 where the Born term and the $O(\alpha_s^3)$ cross section are plotted as functions of $\eta_{\text{cut}}$ for the $q\bar{q}$ and $gg$ channels, where $\eta = (s - 4m^2)/4m^2$ is the variable into which we have incorporated a cut in our programs for the cross sections. The cross sections rise sharply for increasing values of $\eta_{\text{cut}}$ between 0.1 and 1 and they reach a plateau at higher values of $\eta_{\text{cut}}$. This indicates that the threshold region is very important and that the region where $s >> 4m^2$ only makes a small contribution to the cross sections. Note that in the last figure as well as throughout the rest of this paper we are assuming that the bottom quark mass is $m_b = 4.75$ GeV/$c^2$.

Next, we discuss the scale dependence of our NLO results. In fig. 4 we show the total $O(\alpha_s^3)$ cross section as a function of the factorization scale for the $q\bar{q}$ and $gg$ channels. We see that as the scale decreases, the NLO cross sections peak at a scale close to half the mass of the bottom quark and then decrease for smaller values of the scale. For the $q\bar{q}$ channel the NLO cross section is relatively flat. The situation is much worse for the $gg$ channel, however, since the peak is very sharp and the scale dependence is much greater. Since the $gg$ channel dominates, this large scale dependence is also reflected in the total cross section. Thus the variation in the NLO cross section for scales between $m/2$ and $2m$ is large.

In fig. 5 we examine the $\mu_0$ dependence of the resummed cross sections for $b$-quark production at HERA-B for the $gg$ and $q\bar{q}$ channels (in the $\overline{\text{MS}}$ and DIS schemes). We also show, for comparison, the $\mu_0$ dependence of $\sigma^{(0)} + \sigma^{(1)} \mid_{\text{app}} + \sigma^{(2)} \mid_{\text{app}}$, where $\sigma^{(1)} \mid_{\text{app}}$ and $\sigma^{(2)} \mid_{\text{app}}$ denote the approximate first and second order corrections, respectively, where only soft gluon contributions are taken into account via the expansion of (2.11). Note that we have imposed the same cut on the phase space of $s_4$ $(s_4 > s_0)$ as for the resummed cross section. The effect of the resummation shows in the difference between the two curves for each channel. At small $\mu_0$, $\sigma^{\text{res}}$ diverges signalling the divergence of the running coupling constant. This is not physical and should be cancelled by an unknown non-perturbative term. There is a region for each channel where the higher-order terms are numerically important. At large values of $\mu_0$ the two lines for each channel are practically the same. For the $q\bar{q}$ channel in the DIS scheme the resummation is successful in the sense that there is a relatively large region of $\mu_0$ where resummation is well behaved before we encounter the divergence. This region is reduced for the $q\bar{q}$ channel in the $\overline{\text{MS}}$ scheme reflecting the differences between (2.6) and (2.9). For the $gg$ channel, however, this region is even smaller.

From these curves we choose what we think are reasonable values for $\mu_0$. We choose $\mu_0 = 0.6$ GeV for the $q\bar{q}$ channel in the DIS scheme ($\mu_0/m \approx 13\%$) and $\mu_0 = 1.7$ GeV for the $gg$ channel ($\mu_0/m \approx 36\%$). The values we chose for the $q\bar{q}$ and $gg$ channels are such that the resummed cross sections are slightly larger than the sums $\sigma^{(0)} + \sigma^{(1)} \mid_{\text{app}} + \sigma^{(2)} \mid_{\text{app}}$. Note that these $\mu_0$ values are not exactly the same as those used in ref. 4, where $\mu_0/m = 10\%$ and $\mu_0/m = 25\%$ for the $q\bar{q}$ and $gg$ channels respectively, which predicted the mass dependence of the top quark cross section. In this reference the $\mu_0$ parameters were again chosen via the criterion that the higher order terms in the perturbation theory should not be too large.

It is illuminating to compare fig. 5 with a corresponding plot for the top quark case (for instance with figs. 12, 13, and 14 in [3], where the top quark mass was taken as 100 GeV). There one can infer that if we take the slightly larger $\mu_0$ values given above there is very little change in the top quark cross section. The reason is that in this case the $gg$ channel makes only a small contribution and the $\mu_0$ dependence in the $q\bar{q}$ channel reflects the small variation of the running coupling constant at a scale $\mu = 100$ GeV. As the running coupling constant varies more rapidly at a scale $\mu = 4.75$ GeV the $\mu_0$ parameters should be taken from measurements at the lower scale and then used in the prediction of the top quark cross section. This emphasizes the importance of the proposed measurement at HERA-B. It is clear from fig. 5 that we cannot choose $\mu_0/m = 25\%$ for the $gg$ channel for bottom quark production but we can choose $\mu_0/m = 36\%$ for the $gg$ channel for top quark production, with very little change in the value of the top quark cross section. Both sets of parameters yield cross sections which are within the error bars of the recent CDF [14] and D0 [15] experimental results for the top quark cross section. Therefore our cut off parameters do have experimental justification. We would also like to point out that an application of the principal value resummation method has been recently completed by Berger and Contopanagos [16] leading to essentially the same mass dependence of the top cross section as reported in ref. 4, which again justifies our choice for $\mu_0$. Finally note that we could just as easily have chosen to work in the $\overline{\text{MS}}$ scheme for both channels by changing $\mu_0$ in the $q\bar{q}$ channel to $\mu_0 \approx 1.3$ GeV. The reason the DIS scheme is preferred is simply because it has a larger radius of convergence.

In fig. 6 we plot the NLO cross section for $\mu = m/2$, $m$, and $2m$, and the resummed cross section with the values of $\mu_0$ that we chose from fig. 5, as a function of the beam momentum for $b$-quark production at fixed-target $pp$ experiments. The width of the NLO band reflects the large scale dependence that we discussed above. The total NLO cross section for $b$-quark production at HERA-B (beam energy 820 GeV) is 28.8 nb for $\mu = m/2$; 9.6 nb for $\mu = m$; and 4.2 nb for $\mu = 2m$. The resummed cross section is 18 nb. This value was calculated with the cut $s_4 > s_0$ while no such cut was imposed on the NLO result. As we know the exact $O(\alpha_s^3)$ result, we can make an even better estimate using the perturbation theory improved cross section defined by

$$\sigma_H^{\text{imp}} = \sigma_H^{\text{res}} + \sigma_H^{(1)} \mid_{\text{exact}} - \sigma_H^{(1)} \mid_{\text{app}}, \qquad (3.6)$$

to exploit the fact that $\sigma_H^{(1)} \mid_{\text{exact}}$ is known and $\sigma_H^{(1)} \mid_{\text{app}}$ is included in $\sigma_H^{\text{res}}$. Therefore we also plot the improved total cross section versus beam momentum in fig. 6 (where we have taken into account the small negative contributions of the $qg$ and $\bar{q}g$ channels). The improved total cross section for $b$-quark production at HERA-B is 19.4 nb.

It is interesting to make an estimate of the theoretical uncertainty in the resummed cross section. For this one would have to determine the non-leading logarithmic factors in the order-by-order cross section. This is a difficult problem because the $gg$ channel dominates so our knowledge of the order-by-order Drell-Yan cross section is not relevant. The problem is presently under investigation [17]. In [3] the logarithms involving the scale parameter $\mu$ were resummed into the factor $\exp(-\eta\gamma_E)$ in (2.11). This involves the assumption that these factors exponentiate, which may not be true. However we will use (2.11) as an estimate. The scale dependence enters into the running coupling constant, the parton densities, as well as the factors involving $\eta$ in (2.11). To get some idea of the $\mu$ dependence of the resummed cross section in (2.11) we have rerun our programs with $\mu = 4m/5$ and $\mu = 5m/4$ and plotted the results in fig. 6. We see that the cross section is still sensitive to the choice of $\mu$, mainly through the exponential in (2.11). As noted above this is probably misleading and the investigation in [17] should yield a more reliable result.

## 4    Conclusions

We have presented NLO and resummed results for the cross section for bottom quark production at HERA-B. It has been shown that the $gg$ channel is dominant and that the threshold region gives the main contribution to the NLO cross section. The resummation of the S+V logarithms produces an enhancement of the NLO results and yields the value of 20 nb at $\sqrt{S} = 39.2$ GeV for $m_b = 4.75$ GeV/$c^2$. The theoretical error on this number is under investigation.

**ACKNOWLEDGEMENTS**

The work in this paper was supported in part under the contract NSF 93-09888.

# List of Figures

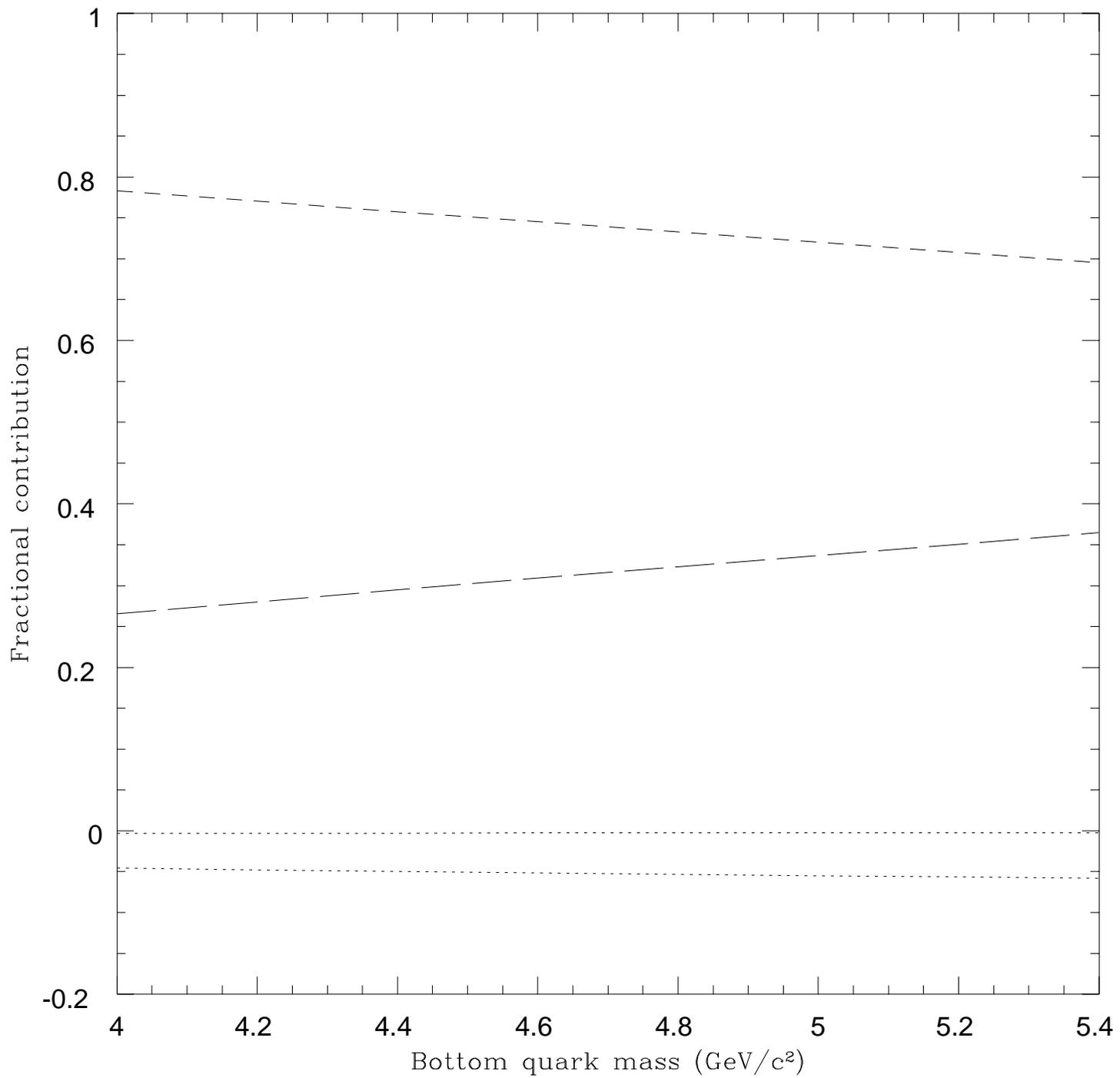

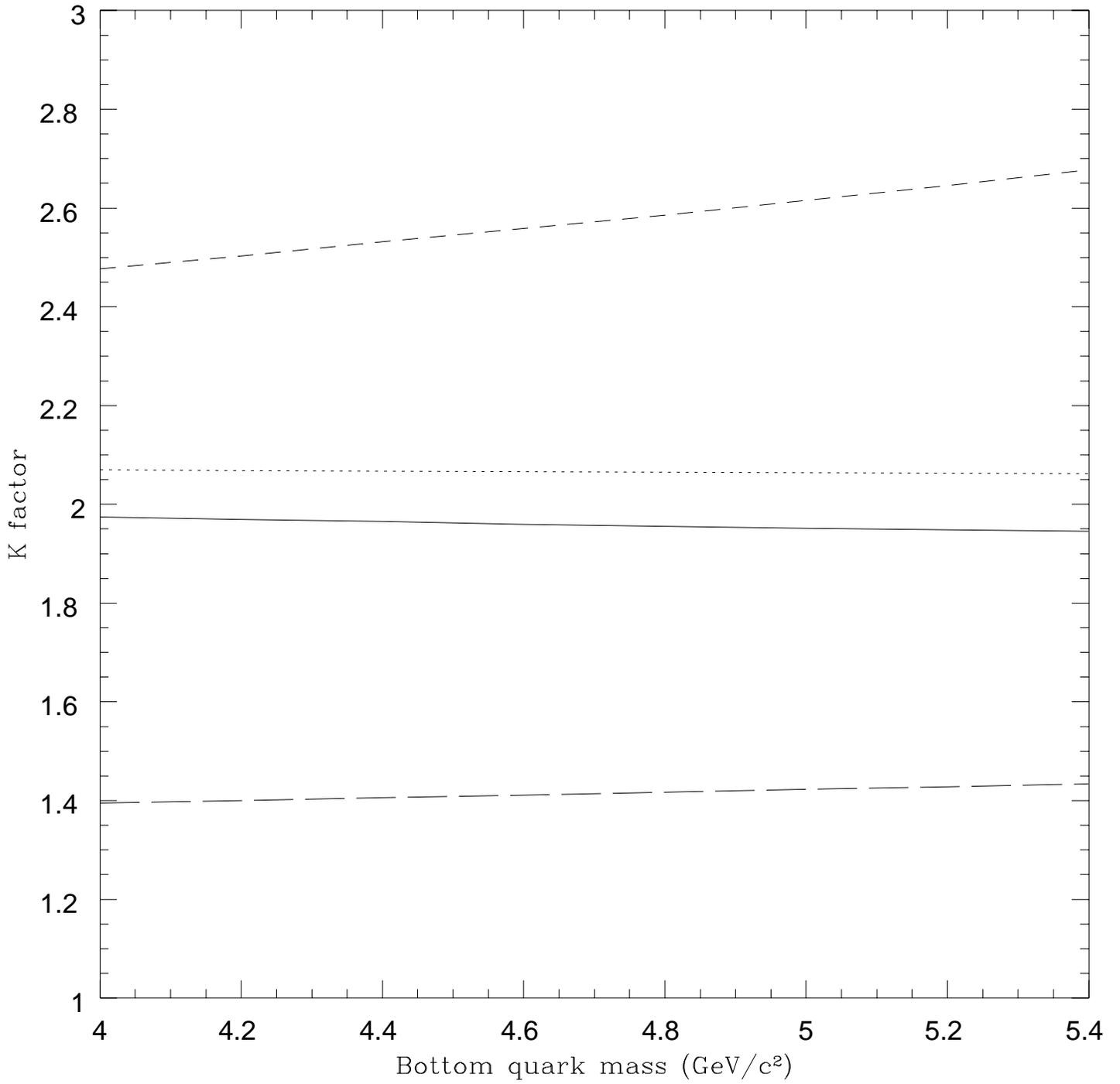

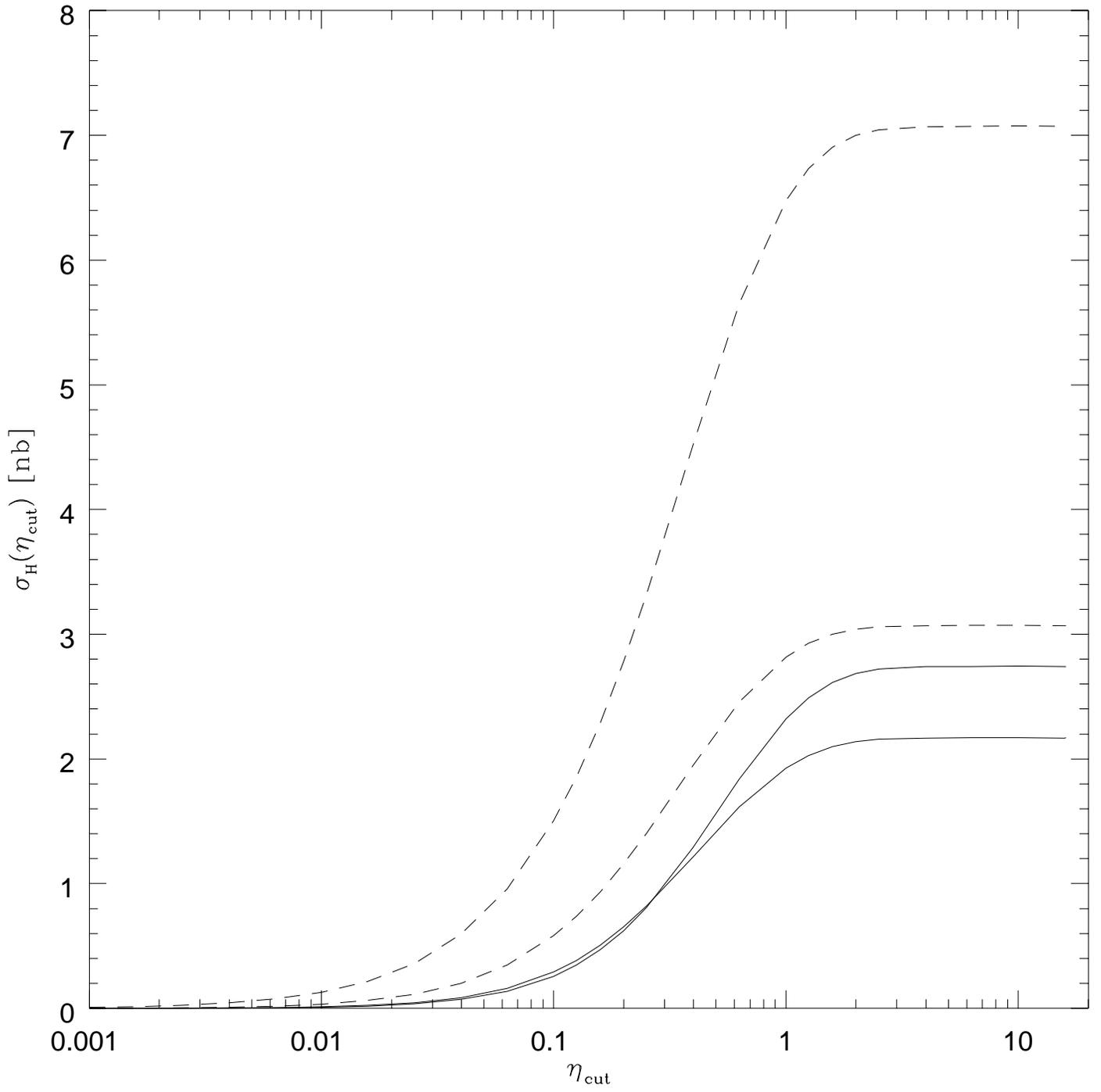

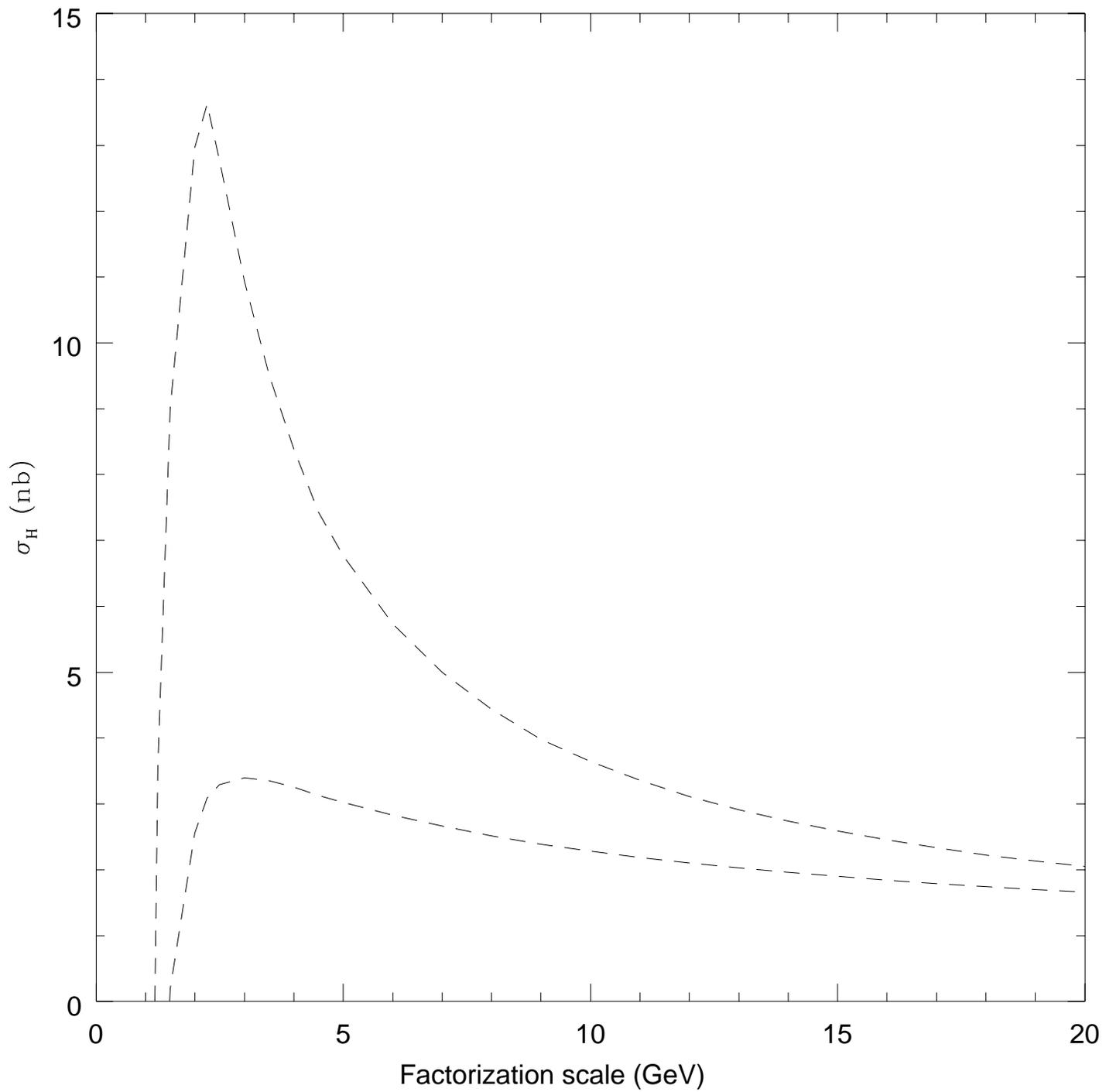

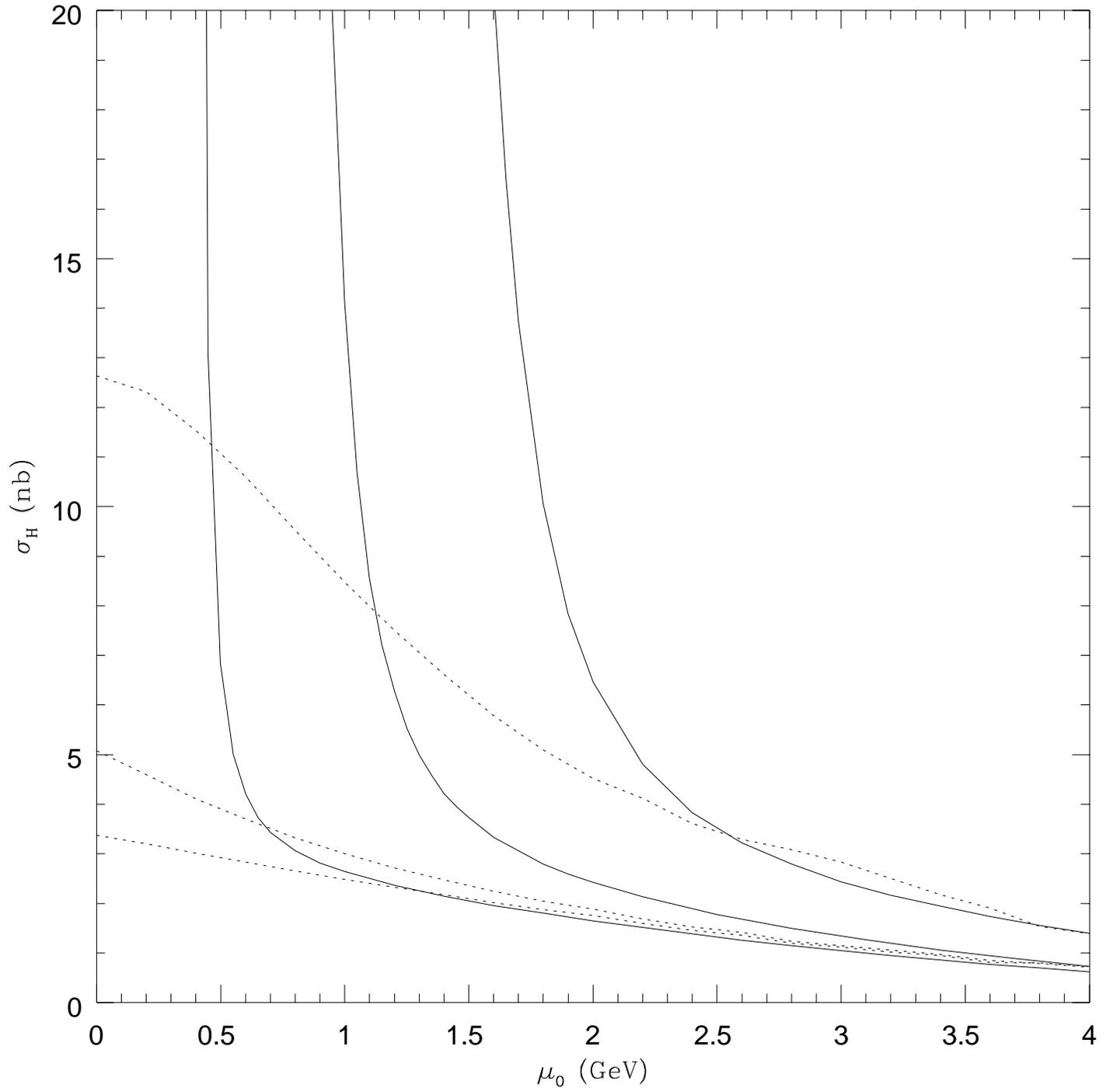

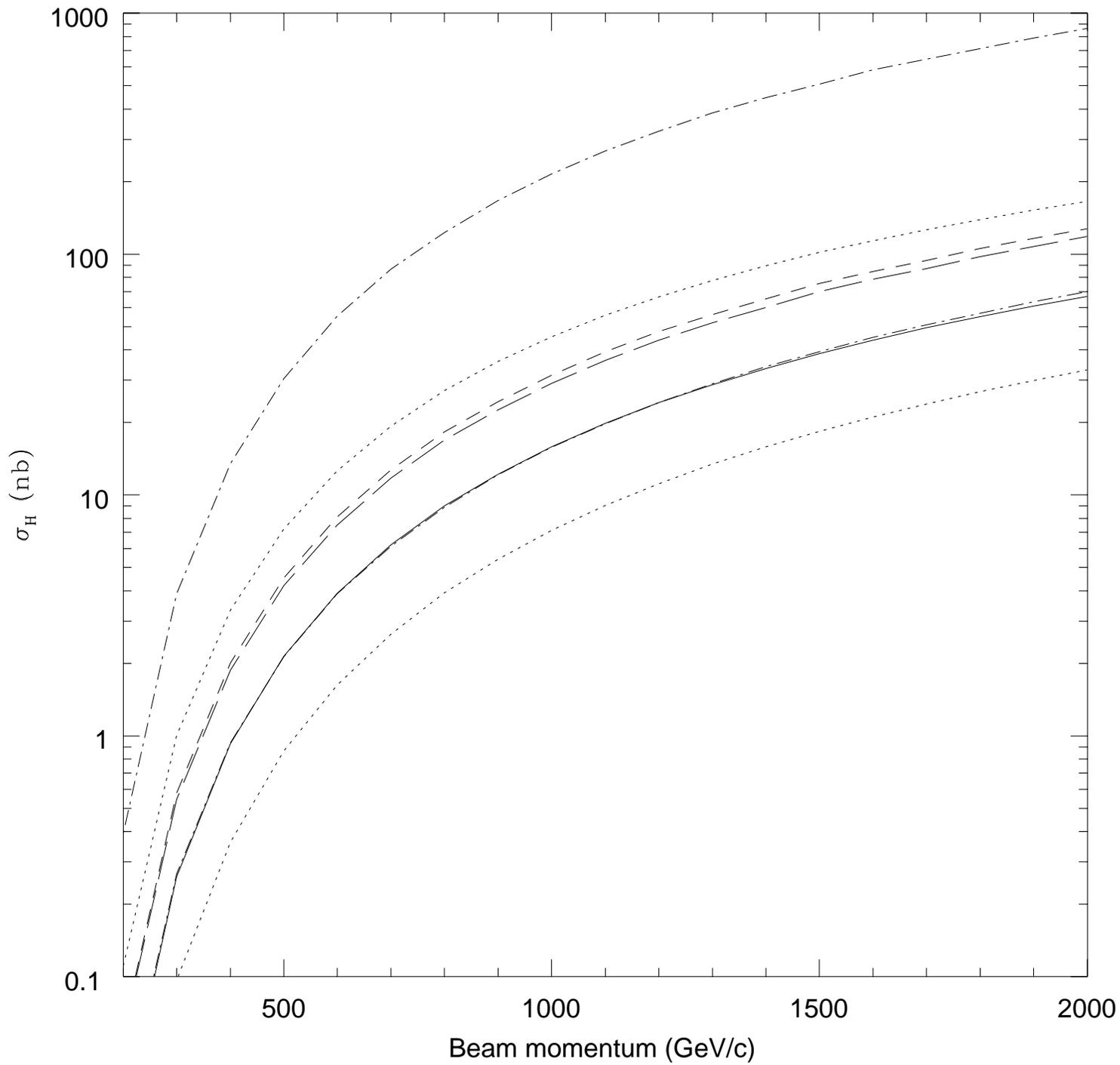